\documentclass[fp,twocolumn]{jpsj3}
\usepackage{txfonts}
\usepackage{color}
\catcode`\@=11
\def\simle{\mathrel{\mathpalette\@versim<}}   % < over \sim
\def\simge{\mathrel{\mathpalette\@versim>}}   % > over \sim
\def\@versim#1#2{\lower2.5pt\vbox{\baselineskip0pt \lineskip-.5pt
  \ialign{$\m@th#1\hfil##\hfil$\crcr#2\crcr\sim\crcr}}}
\catcode`\@=12

\title{Impact of Dzyaloshinsky-Moriya Interactions and Tilts of the $g$ Tensors on the Magnetization Process of a Spherical Kagom\'{e} Cluster in \{W$_{72}$V$_{30}$\}}
\author{Yoshiyuki {\sc Fukumoto}$^1$\thanks{yfuku@rs.tus.ac.jp}, Yuto  {\sc Yokoyama}$^1$, and Hiroki {\sc Nakano}$^2$}
\inst{$^1$Department of Physics, Faculty of Science and Technology, Tokyo University of Science, Noda, Chiba 278-8510\\
$^2$Graduate School of Material Science, University of Hyogo, Ako-gun, Hyogo 678-1297}
\recdate{\hspace{3cm}}
\abst{
In order to clarify why the experimental magnetization curve of the spherical kagom\'{e} cluster in \{W$_{72}$V$_{30}$\} at 0.5 K shows no sign of staircase behavior up to 50 T,
we study the effects of Dzyaloshinsky-Moriya (DM) interactions and tilts of the $g$ tensors, both of which lead to the breaking of the total-$S^z$ conservation, by using the exact diagonalization method.
It is found that the $D$ vector component parallel to the radiation direction of the polyhedron cancels out the staircase in a low magnetic field region efficiently.
The tilts of the $g$ tensors are inherent to systems defined on the polyhedrons and lead to induced magnetic fields varying site by site.
This induced magnetic field affects the magnetization only at high magnetic fields.
We also discuss two existing experimental results on the basis of our calculated results.
}

\begin{document}
\maketitle

%%%%%%%%%%%%%%%%%%%%%%%%%%%%%%%%%%%%%%%%%%%%%%%%%%%%%%%%%%%%%%%%%%%%%%%%%%%%%%%%%%%%%%
\section{Introduction}\label{sec:1}
%%%%%%%%%%%%%%%%%%%%%%%%%%%%%%%%%%%%%%%%%%%%%%%%%%%%%%%%%%%%%%%%%%%%%%%%%%%%%%%%%%%%%%

The study of resonating valence bond (RVB) states is one of central issues in frustrated quantum spin systems,\cite{Anderson1973}
and over the past few decades, a considerable number of studies have been made on the spin-1/2 Heisenberg antiferromagnets in 
%the kagome lattice,
%\cite{Mambrini2000,Yan2011,Depenbrock2012,Lu2011,Nishimoto2013}
%which is a network of corner-sharing triangles.
the kagom\'{e} lattice\cite{Waldtmann1998,Mambrini2000,Hida2001,HN_TSakai_ramp_2010,Nakano2011_2,
HN_kgm_dist,HN_TSakai_kgm_1_3,HN_TSakai_kgm_S,
Yan2011,Depenbrock2012,Lu2011,Nishimoto2013},
which is a network of corner-sharing triangles, 
as well as on the spin-1/2 triangular-lattice Heisenberg
antiferromagnets,\cite{Huse_Elser,Jolicour_LGuillou,Singh_Huse,
Bernu1992,Cubukov1992,Bernu1994,Weihong1999,Richter_lecture2004,
Yunoki2006,Starykh2007,Heidarian2009,Sakai_HN_PRBR,
Harada2012,s1tri_LRO,DYamamoto2013,Starykh_review2015,
HNakano_TSakai_JPSJ2017ferri,
HNakano_TSakai_JPSJ2017plateau} which include
a network of edge-sharing triangles.
In this paper, we study a zero-dimensional counterpart of the kagom\'{e} lattice.
Icosidodecahedron is a network of corner-sharing triangles, and can be considered as a kagom\'{e} on a sphere \cite{Mila2008,Kunisada2008}.
Spin-1/2 icosidodecahedrons, or spherical kagom\'{e} clusters,  are realized in \{W$_{72}$V$_{30}$\} and \{Mo$_{72}$V$_{30}$\}\cite{Muller2005,Botar2005,Todae2009,Li2009}.
The possible relevance of these spherical kagom\'{e} clusters to RVB physics is of great interest\cite{Kunisada2014,Kunisada2015}.

\begin{figure}[b]
\begin{centering}
\includegraphics[width=0.85\linewidth]{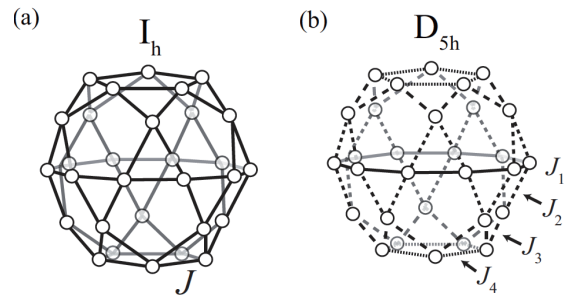}
\caption{Schematic illustrations of (a) $I_{\rm h}$ model and (b) $D_{\rm 5h}$ model.
\label{fig:1}}
\end{centering}
\end{figure}

The main Hamiltonians for these spherical kagom\'{e} clusters are
\begin{equation}
   {\cal H}=\sum_{\langle i,j \rangle}J_{i,j}\mib{S}_i\cdot\mib{S}_j,
   \label{eq:1}
\end{equation}
where $\langle i,j \rangle$ and $J_{i,j}$ denote the nearest neighbors and exchange couplings, respectively.
When $J_{i,j}=J$ for all $\langle i,j \rangle$, the model has $I_{\rm h}$ symmetry and will be referred to as an $I_{\rm h}$ model (see Fig.~\ref{fig:1}(a)).
The symmetry of both \{W$_{72}$V$_{30}$\} and \{Mo$_{72}$V$_{30}$\} is not $I_{\rm h}$ exactly.
Both the materials are more or less distorted and have $D_{\rm{5h}}$ symmetry, as shown in Fig.~\ref{fig:1}(b).
We are especially interested in the $I_{\rm h}$ model from the perspective of RVB physics, because the distortion tends to eliminate its characteristic behavior\cite{Kunisada2014}.

%----------------------------------------------------
%----------------------------------------------------

%Mo72V30について
In 2005, Mo$_{72}$V$_{30}$ was synthesized independently by M\"{u}ller {\it et al.} and Botar {\it et al.},
and the susceptibility measurement indicated a singlet ground state \cite{Muller2005,Botar2005}. 
It was found that the high-temperature susceptibility was reproduced by the $I_{\rm h}$ model with $J=245$ K, but the low-temperature part was not.\cite{Kunisada2008}
As the temperature decreases, the calculated susceptibility of the $I_{\rm h}$ model vanishes much faster than the experimental susceptibility.
In particular, the experimental spin gap is about one-fourth of that of the $I_{\rm h}$ model.
In order to resolve this discrepancy, the distortion in Mo$_{72}$V$_{30}$ was taken into account,\cite{Kunisada2014}
and a distorted model with the $D_{\rm{5h}}$ symmetry was introduced (see Fig.~\ref{fig:1}(b)).
If distortion exists, then each triangle unit becomes an isosceles triangle, which tends to stabilize the ferrimagnetic phase with a total spin of $S\neq 0$.
By adequately choosing four exchange parameters in the $D_{\rm{5h}}$ model , it is possible to reproduce the experimental susceptibility of Mo$_{72}$V$_{30}$.
The small spin gap of Mo$_{72}$V$_{30}$ can be interpreted to indicate the system being in close proximity to the ferrimagnetic phase \cite{Kunisada2014}.

%W72V30について
In 2009, Todae {\it et al.} synthesized W$_{72}$V$_{30}$ and measured its magnetic susceptibility \cite{Todae2009}. 
This material was independently synthesized by Li {\it et al.}, who also carried out a high-energy Raman scattering experiment \cite{Li2009}. 
Schnack studied the thermodynamics of the $I_{\rm h}$ model and suggested, via the analysis of theoretical and experimental susceptibilities, 
that W$_{72}$V$_{30}$ is well described by the $I_{\rm h}$ model \cite{Schnack2010}.
However, the measurement by Todae {\it{et al.}} was not carried out at very low temperatures, where the magnetic susceptibility
shows activation-type behavior, and therefore, it gave no information on the spin-gap energy.
The information on spin-gap energy is crucial to check whether W$_{72}$V$_{30}$ is described by the $I_{\rm h}$ model or not \cite{Kunisada2014}.

%W72V30の磁化測定
Subsequently, magnetization experiments were carried out, in which
the measurement temperature was 0.5 K and the maximum magnetic field was 50 T \cite{Schnack2013,Nojiri2015}.
Surprisingly, the experimental magnetization processes show no staircase behavior, 
although the $I_{\rm h}$ model in (1) shows three steps within the applied magnetic field region.
Schnack {\it et al.} reported that the origin of the luck of the staircase behavior was ascribed to a broad distribution of the nearest-neighbor exchange interactions,
where the width of the distribution was taken to be about 30\% of its average value \cite{Schnack2013}.

In this paper, we study the other two mechanisms that cancel out the staircase behavior in the magnetization process.
The symmetry of icosidodecahedron is not high and antisymmetric Dzyaloshinsky-Moriya (DM) interactions should be present \cite{Hasegawa2004}.
For a spin cluster defined on a polyhedron, the Zeeman interaction is naturally accompanied by an site-dependent induced magnetic field component, as described below.
We first note that the directions of principal axes of the $g$-tensor change site by site in a polyhedron.
It is also natural to expect that the $g$ factor along the radiation direction of the polyhedron is different from that in the tangent plane.
These factors cause the induced magnetic fields to vary site by site.
Both the DM interaction and the site-dependent magnetic field break the conservation of magnetization along the direction of the applied magnetic field,
and can be possible sources for the luck of the staircase behavior in the magnetization process.

We use the exact diagonalization method to study the effects of the DM interaction and the site-dependent magnetic field on the magnetization process.
From the result, we find that the staircase behavior at low magnetic fields, corresponding to the experiment, is eliminated by the modest strength of DM interaction,
if the DM vector contains a certain amount of radiation components.
The site-dependent induced magnetic field due to the tilts of the $g$ tensors affects the magnetization only at high magnetic fields.

This paper is organized as follows.
The DM interaction and induced magnetic field are formulated in \S2.
In \S3, we present our results of exact diagonalization calculations and discuss two existing experimental results.
In \S4, we summarize the results obtained in this study.

%%%%%%%%%%%%%%%%%%%%%%%%%%%%%%%%%%%%%%%%%%%%%%%%%%%%%%%%%%%%%%%%%%%%%%%%%%%%%%%%%%%%%%
\section{Model}\label{sec:2}
%%%%%%%%%%%%%%%%%%%%%%%%%%%%%%%%%%%%%%%%%%%%%%%%%%%%%%%%%%%%%%%%%%%%%%%%%%%%%%%%%%%%%%

\subsection{DM interaction}

\begin{figure}[h]
\begin{center}
\includegraphics[width=.8\linewidth]{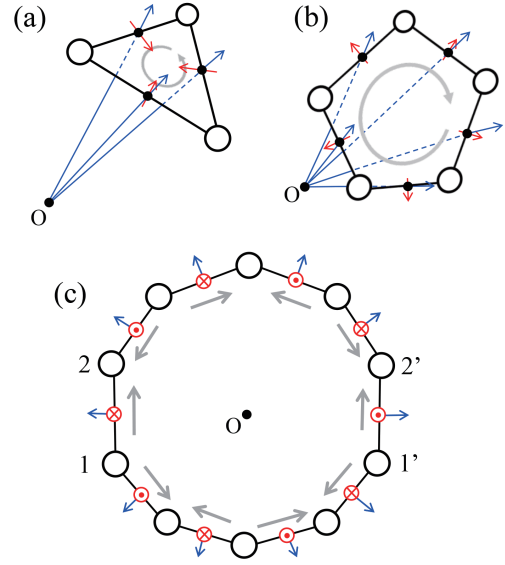}
\caption{Schematic representation of $\mib{e}_{\rm r}^{(i,j)}$ and $\mib{e}_{\rm p}^{(i,j)}$
for (a) triangle and (b) pentagon on the surface of the icosidodecahedron,
and for (c) decagon on the equatorial plane, where $O$ is the center of the icosidodecahedron.
The gray arrows indicate the directions of bonds.
\label{fig:2}}
\end{center}
\end{figure}

We start by defining the direction of a bond $\langle i,j \rangle$.
As shown in Fig.~\ref{fig:2}(a), we order the three sites in each triangle anticlockwise from the outside of an icosidodecahedron.
In the notation $\langle i,j \rangle$, we assume that site $i$ ($j$) is the prior (subsequent) site.
Denoting the position vector of site $i$ ($j$) as $\mib{r}_i$ ($\mib{r}_j$), where the center of the icosidodecahedron is chosen as the origin $O$, 
we introduce the following two unit vectors,
\begin{equation}
   \mib{e}_{\rm r}^{(i,j)}=\frac{\mib{r}_i+\mib{r}_j}{|\mib{r}_i+\mib{r}_j|},\;\mbox{and}\;\;
   \mib{e}_{\rm p}^{(i,j)}=\frac{\mib{r}_i\times\mib{r}_j}{|\mib{r}_i\times\mib{r}_j|},
\end{equation}
which form the basis vectors for the plane normal to the bond direction $\mib{r}_j-\mib{r}_i$.
In Fig.~\ref{fig:2}(a), the blue and red arrows, respectively, represent $\mib{e}_{\rm r}^{(i,j)}$ and $\mib{e}_{\rm p}^{(i,j)}$.
We call the direction of $\mib{e}_{\rm r}^{(i,j)}$ ($\mib{e}_{\rm p}^{(i,j)}$) as the ``radiation (perpendicular) direction" of the bond $\langle i,j \rangle$.
For a triangle, we have three sets of basis vectors, as shown in Fig.~\ref{fig:2}(a).
Considering the $C_3$ rotation around the symmetry axis of the triangle, 
we can easily observe that the three sets are replaced by each other due to the $C_3$ rotation operation.
In the same way, for the pentagon shown in Fig.~\ref{fig:2}(b), the five sets of basis vectors are replaced by each other due to the $C_5$ rotation operation around the symmetry axis of the pentagon.

%----------------------------------------------------
%----------------------------------------------------

Several magnets with pentagonal structures
have attracted considerable attention\cite{Rousochatzakis_PRB2012_
104415,
HNakano_Cairo_lt,Isoda_Cairo_full,Yamaguchi_SR2015} due to the viewpoint 
that the frustration effect produced by pentagons is 
different from that produced by triangles. 
A pentagon in the present system is surrounded by triangles, which
are in an edge-sharing relationship with the pentagons. 
Some systems including polygons surrounded by edge-sharing triangles 
have been investigated. 
In the case of distorted kagom\'{e} lattice, the surrounded polygon is 
hexagon\cite{Hida2001,HN_kgm_dist,HN_TSakai_kgm_1_3},and in the case of square-kagom\'{e} lattice, the surrounded polygon is 
square\cite{Siddharthan_PRB2001_014417,Tomczak_JPA2003_5399,
Richter_CMP2009_507,shuriken_lett,shuriken_dist}. 
These lattices are created by a tiling of polygons in a flat plane.
Since such tiling of pentagons in a flat plane without any voids
is impossible, the present system forms a sphere.
A comparison among the systems on these lattices is 
another interesting future issue. 

Let us return to the DM interactions in the present system.
We write the DM interaction as
\begin{equation}
   {\cal H}_{\rm{DM}}=\sum_{\langle i,j \rangle}\mib{D}_{i,j}\cdot(\mib{S}_i\times\mib{S}_j),
\end{equation}
which is invariant under symmetry operations of $I_{\rm h}$.
We now consider the symmetry to determine the direction of $\mib{D}_{i,j}$.
For a bond $\langle i,j \rangle$, there exists a mirror plane perpendicular to  the bond direction $\mib{r}_j-\mib{r}_i$.
The invariance of the Hamiltonian under the mirror operation leads to $\mib{D}_{i,j}$ lying in the mirror plane.
Using $\{\mib{e}_{\rm p}^{(i,j)}, \mib{e}_{\rm r}^{(i,j)}\}$ as the basis vectors of the mirror plane, we can write
\begin{equation}
   \mib{D}_{i,j}=D_{\rm p}^{(i,j)}\mib{e}_{\rm p}^{(i,j)}+D_{\rm r}^{(i,j)}\mib{e}_{\rm r}^{(i,j)}.
\end{equation}
Next, we use the invariance under the $C_3$ and $C_5$ rotations around the symmetry axes of both the triangle and the pentagon.
Noting that $\mib{e}_{\rm p}$s and $\mib{e}_{\rm r}$s are invariant under these rotations,
this invariance condition holds if $D_{\rm r}^{(i,j)}$ and $D_{\rm p}^{(i,j)}$ are common for all bonds.
Thus, we get the final expression
\begin{equation}
   {\cal H}_{\rm{DM}}=d\sum_{\langle i,j \rangle}(\cos\theta\;\mib{e}_{\rm p}^{(i,j)}+\sin\theta\;\mib{e}_{\rm r}^{(i,j)})\cdot(\mib{S}_i\times\mib{S}_j),
\label{eq:5}   
\end{equation}
where $d$ represents the strength of the $D$ vector and $\theta$ is the angle defining its direction.

The symmetry considerations give no further information on the direction of the $D$ vector.
It is instructive to look into the space inversion operation $C_{\rm i}$, where the inversion center is the origin $O$, i.e., the center of the icosidodecahedron.
In Fig. 2(c), we show a decagon on the equatorial plane of the icosidodecahedron.
We can refer to a preceding study on nanoscale iron rings for more information about the DM interactions in cluster magnets.\cite{Nakano2002}
Defining the site indexes 1, 2, 1$^{\prime}$, and 2$^{\prime}$ as in Fig.~\ref{fig:2}(c), $C_{\rm i}$  relates 1 and 2 to 2$^{\prime}$ and 1$^{\prime}$, respectively.
Noting that the spin operator is an axial vector, we have $(S_1^x,S_1^y,S_1^z)\rightarrow(S_{2^{\prime}}^x,S_{2^{\prime}}^y,S_{2^{\prime}}^z)$ and
$(S_2^x,S_2^y,S_2^z)\rightarrow(S_{1^{\prime}}^x,S_{1^{\prime}}^y,S_{1^{\prime}}^z)$ by $C_{\rm i}$, and thus
\begin{equation}
   \mib{D}_{1,2}\cdot(\mib{S}_1\times\mib{S}_2)\rightarrow -\mib{D}_{1,2}\cdot(\mib{S}_{1^{\prime}}\times\mib{S}_{2^{\prime}}).
\end{equation}
The invariance condition demands $\mib{D}_{1^{\prime},2^{\prime}}=-\mib{D}_{1,2}$.
However, this relation holds already in eq. (\ref{eq:5}).

\subsection{Zeeman term}

We start by introducing a global coordinate system, whose origin is the center of the icosidodecahedron,
and then denoting the unit vectors in the $x$, $y$, $z$ directions as $\mib{e}_1$, $\mib{e}_2$, $\mib{e}_3$.
We set the position vector $\mib{r}_i$ as
\begin{equation}
   \mib{r}_i=x_i\mib{e}_1+y_i\mib{e}_2+z_i\mib{e}_3.
\end{equation}
Next, we introduce a local coordinate system concerned with site $i$.
We define $\mib{e}_1^{(i)}$ and $\mib{e}_2^{(i)}$ as the unit vectors in the tangent plane at the position $\mib{r}_i$,
and the unit vector in the radiation direction as
\begin{equation}
   \mib{e}_3^{(i)}=\frac{\mib{r}_i}{|\mib{r}_i|}.
\end{equation}
The Zeeman interaction for the whole system is given by
\begin{equation}
   {\cal H}_{\rm{Zemman}}=\sum_i {\cal H}_{\rm{Zemman}}^{(i)},
\end{equation}
where
\begin{equation}
   {\cal H}_{\rm{Zemman}}^{(i)}=-\mib{H}\hat{g}_i\mib{S}_i.
\end{equation}
It is naturally expected that the $g$ tensor, $\hat{g}_i$,  is in a diagonal form, if the above-mentioned local coordinate system is used,
\begin{eqnarray}
   {\cal H}_{\rm{Zemman}}^{(i)}&&\hspace{-8mm}=-[\mib{e}_1^{(i)}\cdot\mib{H},\mib{e}_2^{(i)}\cdot\mib{H},\mib{e}_3^{(i)}\cdot\mib{H}]
   \nonumber\\&&\hspace{10mm}
      \left[\begin{array}{ccc}g_{\perp} & 0 & 0\\0& g_{\perp} & 0\\ 0 & 0 & g_{\parallel}\end{array}\right]
      \left[\begin{array}{c}\mib{e}_1^{(i)}\cdot\mib{S}_i \\\mib{e}_2^{(i)}\cdot\mib{S}_i\\\mib{e}_3^{(i)}\cdot\mib{S}_i\end{array}\right]
   \nonumber\\
   &&\hspace{-8mm}=-g_{\perp}\mib{H}\cdot\mib{S}_i-(g_{\parallel}-g_{\perp})(\mib{e}_3^{(i)}\cdot\mib{H})(\mib{e}_3^{(i)}\cdot\mib{S}_i)
   \nonumber\\
   &&\hspace{-8mm}\equiv -g_{\perp}\left(\mib{H}+\delta\mib{H}_i\right)\cdot\mib{S}_i,
\end{eqnarray}
where
\begin{equation}
   \delta\mib{H}_i=\frac{g_{\parallel}-g_{\perp}}{g_{\perp}}\frac{(\mib{H}\cdot\mib{r}_i)\mib{r}_i}{|\mib{r}_i|^{2}}
\label{eq:12}   
\end{equation}
is a site-dependent induced field in the radiation direction.
The site-dependent magnetic fields break the conservation of the total $S^z$.
It should be stressed that the existence of such additional fields is inherent to systems that are defined on polyhedrons.

%%%%%%%%%%%%%%%%%%%%%%%%%%%%%%%%%%%%%%%%%%%%%%%%%%%%%%%%%%%%%%%%%%%%%%%%%%%%%%%%%%%%%%
\section{Calculation Methods and Results}\label{sec:3}
%%%%%%%%%%%%%%%%%%%%%%%%%%%%%%%%%%%%%%%%%%%%%%%%%%%%%%%%%%%%%%%%%%%%%%%%%%%%%%%%%%%%%%

\subsection{Calculation methods}

In the magnetization measurements, polycrystalline samples of \{W$_{72}$V$_{30}$\} were used.\cite{Schnack2013,Nojiri2015}
The orientations of magnetic clusters in such samples are distributed randomly with respect to the magnetic-field direction.
In order to reproduce this situation in our theoretical calculations, we prepare a sample set of randomly distributed magnetic fields, 
\{$\mib{H}_1,\mib{H}_2,\cdots,\mib{H}_{N_{\rm s}}$\}, where $|\mib{H}_l|=H$ for all $l$.
Then, we apply $\mib{H}_l$ on a magnetic cluster and calculate the induced magnetization, $M_l$, parallel to the field.
The observed magnetization is estimated from
\begin{equation}
   M(H)=\frac{1}{N_{\rm s}}\sum_{l=1}^{N_{\rm s}}M_l.
\end{equation}
Our calculated results shown below are accompanied by error bars resulting from the statistical error.
In general, for the magnetization steps in a system that has only exchange interactions, we need to use larger values of $N_{\rm s}$ around the magnetic fields .
In the following calculated results, the maximum sampling size is $N_{\rm s}\sim 50$.

As the present system does not conserve the total $S^z$, the dimension of the Hilbert space is $2^{30}\simeq 1.07\times 10^9$.
In our exact diagonalization calculation via the Lanczos method, we work with the standard Ising basis, and therefore, the matrix elements of DM and Zeeman terms are complex.
We need to prepare three complex vectors with a dimension of $\sim 10^9$, whose memory usage is 48 GB.
Our numerical calculations were performed on the SGI Altix ICE 8400EX at the Supercomputer Center, Institute for Solid State Physics, University of Tokyo,
using OpenMP parallelization with up to 24 cores.

\subsection{Calculation results}

\subsubsection{Case of $d=0$}

Here, we study the effect of tilts of the $g$ tensors.
We set $d=0$, and introduce a parameter,
\begin{equation}
   R_g=\frac{g_{\parallel}-g_{\perp}}{g_{\perp}},
\end{equation}
which determines the strength of the induced field in (\ref{eq:12}).
The calculated results for $R_g=\pm 0.1$, together with that  for $R_g=0$ are shown in Fig. 3.
In Fig. 3, we find that the effect of $R_g\neq 0$ appears only at high magnetic fields.
Therefore, the tilts of the $g$ tensors is not the reason for the absence of staircase behavior up to 50 T in the experimental observations.

%----------------------------------------------------
\begin{figure}[h]
\begin{center}
\includegraphics[width=.85\linewidth]{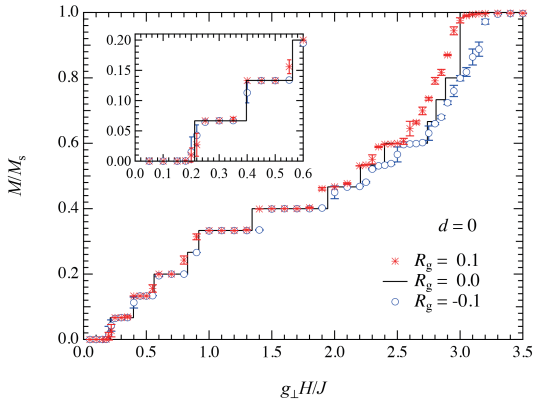}
\end{center}
\caption{Magnetization process for the entire magnetic field region, when the tilts of the $g$ tensors are taken into account.}
\label{fig:3}
\end{figure}
%----------------------------------------------------

\subsubsection{Case of $R_g=0$}

Here, we consider the case of $g_{\parallel}=g_{\perp}\equiv g$, which means no induced field, i.e. $\delta\mib{H}_i=0$.

After setting the strength of the $D$ vector to $d=0.1J$, we start by studying the direction dependence of the low-field magnetization process.
The calculated results are shown in Fig. 4,
\textcolor{black}{
together with a line with the slope corresponding to the peak value of the magnetic susceptibility, $\chi_{\rm{peak}}$, for the $I_{\rm h}$ model without DM interactions.\cite{Kunisada2014}
The spin gap at $d=0$ is corresponding to $gH/J=0.218$.
In the magnetic field region $0\leq gH/J\leq 0.6$, which is comparable to the experimental region, the magnetization process for $d=0.0$ contains four steps.
}
There is a pronounced hint of staircase behavior when the radiation direction component is small, $\theta\simeq 0$ and $\pi$.
Conversely, it is noticed that the radiation direction component tends to eliminate the staircase behavior.
\textcolor{black}{
If the radiation direction component is not very small, then magnetization more than the spin-gap field does not depends on $\theta$ vary much and the slopes are close to $\chi_{\rm{peak}}$.
On the other hand, magnetization lower than the spin-gap field depends on $\theta$, which is quite natural because the DM interaction determines the value of zero-temperature susceptibility.
}
On comparing $\theta= 0.5\pi$ and $1.5\pi$, the latter has a weaker cusp 
\textcolor{black}{
around the spin-gap field 
}
and seems to reproduce the experimental result better.

%----------------------------------------------------
\begin{figure}[h]
\begin{center}
\includegraphics[width=.85\linewidth]{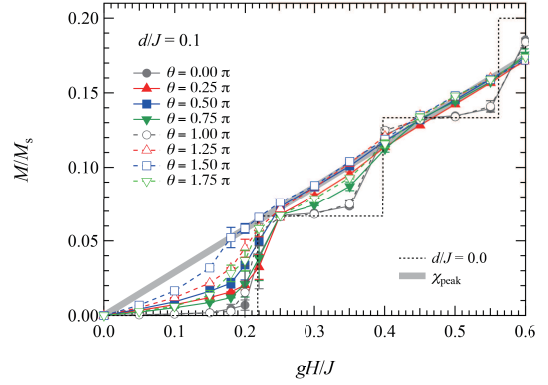}
\end{center}
\caption{Low-field magnetization process for various values of $\theta$.
\textcolor{black}{
The slope of the gray line corresponds to $\chi_{\rm{peak}}$ for the $I_{\rm h}$ model with $d=0$.
}
}
\label{fig:4}
\end{figure}
%----------------------------------------------------

Fig.~5 shows the dependence of low-field magnetization processes on $d$ up to $0.20J$, when the direction is fixed to be $\theta=1.5\pi$.
As $d$ increases, the staircase behavior tends to fade away.
\textcolor{black}{
It is found, as in Fig.~4, that magnetization more than the spin-gap field does not depends on $d$ vary much and the slopes are close to $\chi_{\rm{peak}}$.
However, the dependence on $d$ is pronounced for magnetization lower than the spin-gap field.
When $d$ increases, the zero temperature susceptibility, $\chi_{T=0}$,  increases and we have $\chi_{T=0}>\chi_{\rm{peak}}$ for $d/J \simge 0.15$.
}

%----------------------------------------------------
\begin{figure}[h]
\begin{center}
\includegraphics[width=.85\linewidth]{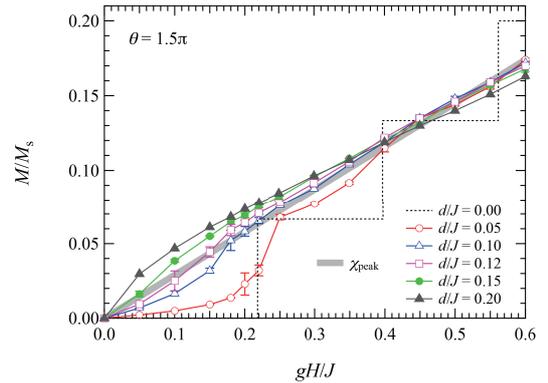}
\caption{Low-field magnetization process for various values of $d$.
}
\end{center}
\label{fig:5}
\end{figure}
%----------------------------------------------------

\subsubsection{The entire magnetization process}

The entire magnetization processes for $R_g=0,\;\pm 0.05$ up to the saturation field are shown in Fig. 6, when $d=0.1J$ and $\theta=1.5\pi$.
We find that the steps below the one-third plateau, $gH/J \simle 1$, are cancelled out by the radiation component of the $D$ vector.
For the intermediate magnetic field region of $1.4<g_{\perp}H/J<2.7$, if $d=0$ and $R_g=0$, 
then there exist wide steps at $M/M_{\rm s}=6/15,\;7/15,\;8/15,\;9/15$.
The signatures of these steps are expected to remain, even if we set modest values of $d$ and $R_g$.
For the high magnetic field region of $2.7<g_{\perp}H/J$, each step is narrow from the beginning. 
These steps are readily smoothed out by modest values of $d$ and $R_g$.

%----------------------------------------------------
\begin{figure}[h]
\begin{center}
\includegraphics[width=.85\linewidth]{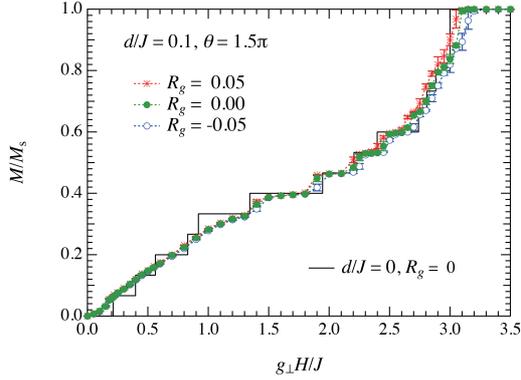}
\end{center}
\caption{Magnetization process for the entire magnetic field region.}
\label{fig:6}
\end{figure}
%----------------------------------------------------

\subsection{Comparison with experimental data}

Here, we use $\theta=1.5\pi$ and $R_{\rm g}=0$ in our theoretical calculations.
In Fig. 7, our calculated magnetization curves are compared with the existing experimental data from two studies.\cite{Schnack2013,Nojiri2015}

The first one was observed at $T=0.5$ K in 2013,\cite{Schnack2013} which exhibits a linear behavior of $M$ with respect to $H$ approximately.
The black line in Fig.~7 represents the linear approximation of their magnetization curve.
We find that all over behaviour of this experimental result is reproduced well by the theoretical calculation for $d=0.1J$, which is shown by the closed squares in Fig. 7.
\textcolor{black}{
However, as mentioned before, the calculated magnetization deviates from the linear approximation below the spin gap field.
As for the sample used in Ref.~\ref{Schnack2013}, it was described that there exist two uncorrected ions, VO$^{2+}$, per a \{W$_{72}$V$_{30}$\} molecule.
It may be plausible that the subtraction procedure of free V$^{4+}$ spins from raw magnetization data washes out such a structure.
In fact, the magnetization curve shown in Fig.~3 in Ref.~\ref{Schnack2013} takes negative values below 5 T.
}

\textcolor{black}{
Another experimental magnetization curve was presented by Nojiri {\it et al.} in 2015,\cite{Nojiri2015} 
where the measurement temperature $T=1.6$ K is somewhat higher than the first one but still much lower than the spin-gap energy $0.218J=21.5$ K.\cite{Kunisada2014}
They treated the number of  free V$^{4+}$ spins as a fitting parameter to decompose their raw magnetization curve into a linear term and Brillouin function.\cite{Kihara2017}
In Fig.~7, the red solid line represents the linear term, which is intrinsic to the spherical kagom\'{e} cluster, and the red dotted line represents the Brillouin function due to impurity spins.
It found that the calculated data at low magnetic fields are on the linear term.
Their procedure seems to be adequate at low magnetic fields below 10 T.
}

%----------------------------------------------------
\begin{figure}[h]
\begin{center}
\includegraphics[width=.85\linewidth]{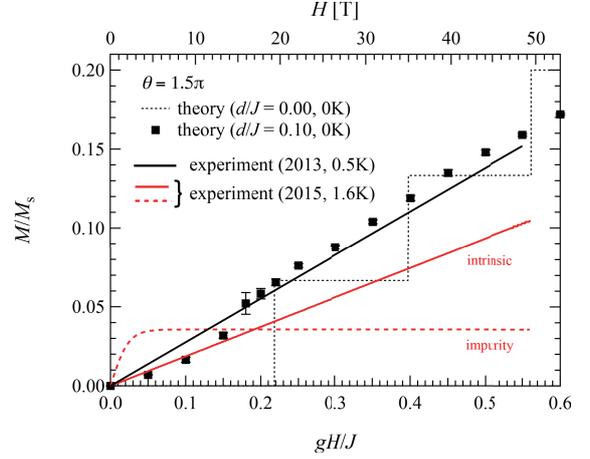}
\end{center}
\caption{Comparison between the theoretical and experimental magnetization curves,
with $J=115$K and $g=1.95$.\cite{Kunisada2014}
The earlier experimental data from 2013 was reported in Ref.~\ref{Schnack2013}, and the latest data from 2015 was reported in Ref.~\ref{Nojiri2015}.

}
\label{fig:7}
\end{figure}
%----------------------------------------------------

%%%%%%%%%%%%%%%%%%%%%%%%%%%%%%%%%%%%%%%%%%%%%%%%%%%%%%%%%%%%%%%%%%
\section{Summary and Future Problems}\label{sec:4}
%%%%%%%%%%%%%%%%%%%%%%%%%%%%%%%%%%%%%%%%%%%%%%%%%%%%%%%%%%%%%%%%%%

We have performed exact diagonalization calculations for the magnetization process of a spherical kagom\'{e} cluster, incorporating the DM interaction and the tilts of the $g$-tensors.
It was found that the low-field magnetization steps are efficiently cancelled out by the DM interactions with the $D$ vector parallel to the radiation direction.
\textcolor{black}{
When the radiation component is not so small, we have the linear magnetization process with slope of $\sim\chi_{\rm{peak}}$ at low magnetic fields above the spin gap field. 
Below the spin gap field, we have  another linear term with slope of $\chi_{T=0}$ which depends on the strength and direction of $D$ vector.
} 
We have compared two existing experimental magnetization curves with our calculated results
\textcolor{black}{
to point out that two different linear magnetization processes reported in Refs.~\ref{Schnack2013} and \ref{Nojiri2015} may, respectively, correspond to $\chi_{\rm{peak}}\times H$
and $\chi_{T=0}\times H$.
} 

The reason why we expected the $I_{\rm h}$ model to be adequate for \{W$_{72}$V$_{30}$\} was 
that the experimental susceptibility down to 5 K reported in Ref.~\ref{Todae2009} was consistent with the theoretical susceptibility of the $I_{\rm h}$ model.\cite{Schnack2010}
However, the experimental data was provided in the form of $\chi T$, and thus it was not clear whether the theoretical and experimental results agree with each other at low temperatures. 
In a later susceptibility experiment in 2013, measurements down to 2.5 K were carried out and the experimental data in the form of $\chi$ were presented.\cite{Schnack2013}
This latest susceptibility data shows a broad peak around 10-20 K and a rapid decrease below about 5 K.
On the other hand, the $I_{\rm h}$ model leads to a rapid decrease below about 10-15 K, and thus does not reproduce the experimental result at low temperatures.
The experimental susceptibility indicates that there are more low-energy magnetic components in \{W$_{72}$V$_{30}$\} than in the $I_{\rm h}$ model,
which seems to be consistent with the existence of DM interactions.
\textcolor{black}{
Also, we note that the susceptibility experiment gives $\chi_{2.5 {\rm K}}/\chi_{\rm{peak}}\simeq 0.7$, which is not contradict to the results shown in Fig.~7.
}

\textcolor{black}{
It is desired to study theoretically how DM interactions affect the magnetic susceptibility quantitatively.
However, DM interactions break the conservation of total spin and get the evaluation of magnetic susceptibility to be more difficult. 
As for kagom\'{e}-lattice antiferromagnets with DM interactions, 
Rigol and Singh calculated magnetic susceptibility up to a 15-site cluster to understand the experimental susceptibility of Herbertsmithite.\cite{Rigol2007}
They calculated all eigenvalues under magnetic fields $H$, and evaluated magnetic susceptibility from the second-order coefficient in $H$ of the free energy.
As for the 30-site spherical kagom\'{e} cluster, the calculation of free energy is possible,\cite{Kunisada2014}
but it is accompanied by statistical error being an serious obstacle for the numerical evaluation of the second-order coefficient.
Instead, we can use the method of thermal pure quantum states to calculate magnetizations at finite temperatures,\cite{Sugiura2012} whose first-order coefficients give the magnetic susceptibility as a function of temperature.
The study along this line is now proceeding.
}

The main concern in the study of spherical kagom\'{e} clusters is the exploration of the characteristic behavior of RVB states.
\textcolor{black}{
From this perspective, it is desired to make experimental studies on specific heat and Raman spectrum, by which we can access low-energy states in the singlet sector directly.\cite{Kunisada2014}
}

\section*{Acknowledgments}

\begin{acknowledgments}
We acknowledge Dr. K. Kihara and Professor H. Nojiri for providing their unpublished data and fruitful discussions.
The authors thank the Supercomputer Center, Institute for Solid State Physics, University of Tokyo for the use of their facilities. 
This work was partly supported by JSPS KAKENHI Grant Numbers 17K05519, 16K05418, 16K05419, and 16H01080 (JPhysics).
\end{acknowledgments}

\end{document}